\documentclass{article}

\usepackage[english]{babel}

\usepackage[letterpaper,top=2cm,bottom=2cm,left=3cm,right=3cm,marginparwidth=1.75cm]{geometry}

\usepackage[most]{tcolorbox}
\usepackage[dvipsnames]{xcolor}
\usepackage{amsmath, amsthm}
\usepackage{amssymb}
\usepackage[normalem]{ulem}

\newtcolorbox{definition}[1][]{
  colback=blue!5!white,
  colframe=blue!75!black,
  fonttitle=\bfseries,
  title={Definition},
  #1
}

\newtcolorbox{theorem}[1][]{
  colback=green!5!white,
  colframe=green!60!black,
  fonttitle=\bfseries,
  title={Theorem},
  #1
}

\newtcolorbox{example}[1][]{
  colback=orange!5!white,
  colframe=orange!75!black,
  fonttitle=\bfseries,
  title={Example},
  #1
}

\newtcolorbox{uniformexample}[1][]{
  colback=orange!5!white,
  colframe=orange!75!black,
  fonttitle=\bfseries,
  title={Uniform Example},
  #1
}

\newtcolorbox{remark}[1][]{
  colback=gray!5!white,
  colframe=gray!60!black,
  fonttitle=\bfseries,
  title={Remark},
  #1
}

\newtcolorbox{important}[1][]{
  colback=red!5!white,
  colframe=red!75!black,
  fonttitle=\bfseries,
  title={!},
  #1
}

\newtcolorbox{rules}[1][]{
  colback=blue!5!white,
  colframe=blue!75!black,
  fonttitle=\bfseries,
  title={3 Rules},
  #1
}

\newtcolorbox{exercise}[1][]{
  colback=violet!5!white,
  colframe=violet!75!black,
  fonttitle=\bfseries,
  title={Exercise},
  #1
}

\usepackage{amsmath}
\usepackage{graphicx}
\usepackage[colorlinks=true, allcolors=blue]{hyperref}
\usepackage[ruled,vlined]{algorithm2e}
\usepackage{algorithmic}
\usepackage{varioref}
\usepackage{cleveref}
\usepackage{longtable}
\RequirePackage[numbers]{natbib}
\usepackage{tikz}
\usetikzlibrary{arrows.meta,positioning,shapes,calc}
\usepackage{array}
\usepackage{colortbl}

\definecolor{headerblue}{RGB}{70,120,180}
\definecolor{missingred}{RGB}{220,80,80}
\definecolor{imputedorange}{RGB}{255,140,80}
\definecolor{predictgreen}{RGB}{120,200,120}

\newcommand{\datatable}[3]{%
    \begin{tabular}{|c|c|c|}
    \hline
    \rowcolor{headerblue}
    \textcolor{white}{\scriptsize\textbf{Age}} & \textcolor{white}{\scriptsize\textbf{Income}} & \textcolor{white}{\scriptsize\textbf{Gender}} \\
    \hline
     #1 \\
    \hline
     #2 \\
    \hline
     #3 \\
    \hline
    \end{tabular}%
}

\newcommand{\datatabletwo}[3]{%
    \begin{tabular}{|c|c|c|}
    \hline
    \rowcolor{headerblue}
    \textcolor{white}{\scriptsize\textbf{Age}} & \textcolor{white}{\scriptsize\textbf{Income}} & \textcolor{white}{\scriptsize\textbf{Gender}} \\
    \hline
    \rowcolor{green!30}
     #1 \\
    \hline
     \rowcolor{green!30}
     #2 \\
    \hline
     #3 \\
    \hline
    \end{tabular}%
}

\newcommand{\datatablethree}[3]{%
    \begin{tabular}{|c|c|c|}
    \hline
    \rowcolor{headerblue}
    \textcolor{white}{\scriptsize\textbf{Age}} & \textcolor{white}{\scriptsize\textbf{Income}} & \textcolor{white}{\scriptsize\textbf{Gender}} \\
    \hline
     #1 \\
    \hline
     \rowcolor{green!30} 
     #2 \\
    \hline
      \rowcolor{green!30} 
     #3 \\
    \hline
    \end{tabular}%
}

\title{A Practical Guide to Modern Imputation}

\author{Jeffrey Näf, \vspace{0.2cm} \\       Research Institute for Statistics and Information Science,\\
        University of Geneva \\
}

\begin{document}
\maketitle

\begin{abstract}
This guide based on recent papers should help researchers avoid some of the most common pitfalls of missing value imputation imputation.
\end{abstract}

\section{Introduction}

Missing values are everywhere, as any reader of this guide is likely aware. One powerful tool in dealing with missing values is imputation. In particular, modern nonparametric imputation tends to deliver (surprisingly) good results in practice, even in very difficult examples. On the other hand, there are serious pitfalls that can render any further analysis heavily biased. Considering recent benchmarks and their recommendations (for instance, \cite{benchmark_jager2021, benchmark_knn_adv1, benchmark_metabolomics1, benchmark_joel2024performance, benchmark_pereira2024imputation, benchmark_alam2023investigation, benchmark_pavelchek2023imputation, benchmark_ge2023simulation,benchmark_seu2022intelligent, benchmark_deforth2024performance, benchmark_poulos2018missing,benchmark_getz2023performance,benchmark_wang2022deep, benchmark_wongkamthong2023comparative, benchmark_miao2022experimental}) suggest that these pitfalls are, unfortunately, very common. We hope that this guide will help to make researchers more aware of these pitfalls and how to counter them.

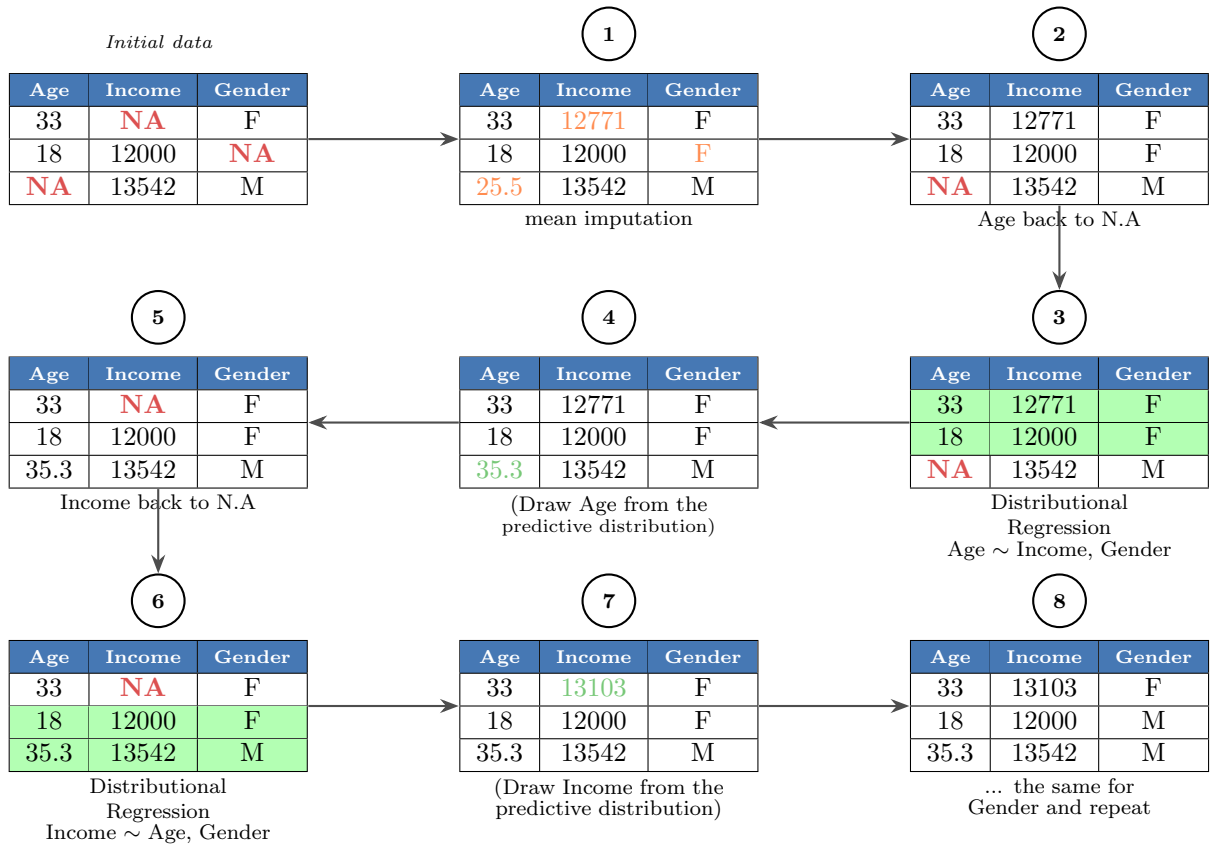
\begin{figure}
    \centering
   \begin{tikzpicture}[
    node distance=0.4cm and 2cm,
    table/.style={inner sep=0pt},
    arrow/.style={-{Stealth[length=2.5mm]}, thick, black!70},
    stepnum/.style={circle, draw=black, fill=white, font=\footnotesize\bfseries, minimum size=7mm, line width=0.8pt},
    label/.style={text width=3.2cm, align=center, font=\scriptsize},
    highlight/.style={circle, line width=1.5pt, minimum size=5mm}
]

\node[table] (t0) {\datatable
    {33 & \textcolor{missingred}{\textbf{\textrm{NA}}} & F}
    {18 & 12000 & \textcolor{missingred}{\textbf{\textrm{NA}}}}
    {\textcolor{missingred}{\textbf{\textrm{NA}}} & 13542 & M}};
\node[above=0.2cm of t0, font=\scriptsize\itshape] {Initial data};

\node[table, right=2cm of t0] (t1) {\datatable
    {33 & \textcolor{imputedorange}{12771} & F}
    {18 & 12000 & \textcolor{imputedorange}{F}}
    {\textcolor{imputedorange}{25.5} & 13542 & M}};
\node[stepnum, above=0.15cm of t1] (s1) {1};
\node[label, below=-0.05cm of t1] {\footnotesize mean imputation};

\node[table, right=2cm of t1] (t2) {\datatable
    {33 & 12771 & F}
    {18 & 12000 & F}
    {\textcolor{missingred}{\textbf{\textrm{NA}}} & 13542 & M}};
\node[stepnum, above=0.15cm of t2] (s2) {2};
\node[label, below=-0.05cm of t2] {\footnotesize Age back to N.A};

\node[table, below=2cm of t2] (t3) {\datatabletwo
    {33 & 12771 & F}
    {18 & 12000 & F}
    {\textcolor{missingred}{\textbf{\textrm{NA}}} & 13542 & M}};
\node[stepnum, above=0.15cm of t3] (s3) {3};
\node[label, below=-0.05cm of t3] {\footnotesize Distributional Regression\\  Age $\sim$ Income, Gender};

\node[table, left=2cm of t3] (t4) {\datatable
    {33 & 12771 & F}
    {18 & 12000 & F}
    {\textcolor{predictgreen}{35.3} & 13542 & M}};
\node[stepnum, above=0.15cm of t4] (s4) {4};
\node[label, below=-0.05cm of t4] { \footnotesize (Draw Age from the\\ \scriptsize predictive distribution)};

\node[table, left=2cm of t4] (t5) {\datatable
    {33 & \textcolor{missingred}{\textbf{\textrm{NA}}} & F}
    {18 & 12000 & F}
    {35.3 & 13542 & M}};
\node[stepnum, above=0.15cm of t5] (s5) {5};
\node[label, below=-0.05cm of t5] {\footnotesize Income back to N.A};

\node[table, below=2cm of t5] (t6) {\datatablethree
    {33 & \textcolor{missingred}{\textbf{\textrm{NA}}} & F}
    {18 & 12000 & F}
    {35.3 & 13542 & M}};
\node[stepnum, above=0.15cm of t6] (s6) {6};
\node[label, below=-0.05cm of t6] {\footnotesize Distributional Regression\\ Income $\sim$ Age, Gender};

\node[table, right=2cm of t6] (t7) {\datatable
    {33 & \textcolor{predictgreen}{13103} & F}
    {18 & 12000 & F}
    {35.3 & 13542 & M}};
\node[stepnum, above=0.15cm of t7] (s7) {7};
\node[label, below=-0.05cm of t7] {\footnotesize (Draw Income from the\\ predictive distribution)};

\node[table, right=2cm of t7] (t8) {\datatable
    {33 & 13103 & F}
    {18 & 12000 & M}
    {35.3 & 13542 & M}};
\node[stepnum, above=0.15cm of t8] (s8) {8};
\node[label, below=-0.05cm of t8] {\footnotesize ... the same for Gender and repeat};

\draw[arrow] (t0) -- (t1);
\draw[arrow] (t1) -- (t2);
\draw[arrow] (t2.south) -- ++(0,-0.6) -| (s3);
\draw[arrow] (t3) -- (t4);
\draw[arrow] (t4) -- (t5);
\draw[arrow] (t5.south) -- ++(0,-0.6) -| (s6);
\draw[arrow] (t6) -- (t7);
\draw[arrow] (t7) -- (t8);







    {Iterate steps 2-8 until convergence (typically 5-10 iterations)};

\end{tikzpicture}
    \caption{Illustration of the mice imputation algorithm, originally from the fantastic video \url{https://www.youtube.com/watch?v=zX-pacwVyvU}, recreated and adapted using Ti\textit{k}Z. Note that in accordance to \textcolor{blue}{Rule 1} in Section \ref{Sec_1}, the mice algorithm in each step learns a (conditional) distribution of one variable given all the others and depending on what method one chooses (linear regression with Gaussian errors, random forest, etc) a different method emerges.}
    \label{fig:miceillustration}
\end{figure}

First, there are a myriad of imputation methods with various flavors; the recent benchmark in \cite{OneBenchmarktorulethemall} collects more than 70 methods. We will illustrate a few methods that can easily be accessed in  \textsf{R}, referenced in Table \ref{tab:methods}, including a few variations of the famous multiple imputation by chained equations (mice) methodology \citep{mice}, illustrated in Figure \ref{fig:miceillustration}. When used to its full potential, it is hard to understate how powerful this iterative approach is. However, the goal of this guide is not necessarily to give a recommendation of what imputation methods to use. Instead, this guide is designed to showcase the aforementioned pitfalls and help applied researchers avoid them. We also present a way to choose the best imputation method for a given task based on \cite{ImputationScores, näf2025Iscore} and discuss a ``challenge'' that can be used to assess how well an imputation method might perform.  


\begin{remark}
    We hope that through lively feedback this guide may grow and improve over time, making it more readable, while its core message remains. In particular, we are happy to obtain feedback from actual research and data science work with missing values and whether the points we make in this article hold up in practice.
\end{remark}

In the following, we draw heavy inspiration from \cite{näf2025good, näf2025Iscore, OneBenchmarktorulethemall} in Sections \ref{Sec_1} and \ref{Sec_2}. Then we discuss a few insights about uncertainty quantification with imputation in Section \ref{Sec_3}. As such, we note that this guide is not free of bias and focuses on issues that we find important in imputation. This might be remedied in new versions of this guide, but will likely not completely vanish. In particular, our guide is centered around iterative imputation as in the mice algorithm, or the well-known knn imputation and focuses on classical i.i.d. tabular data sets of moderate size (i.e. large $n$ small $p$). In such settings at least, it seems the iterative/mice methods perform extremely well, as explored extensively in \cite{OneBenchmarktorulethemall}.

\begin{important}
    Machine learning papers in conferences such as NeurIPS (e.g. \cite{GAIN} and many follow up papers), as well as a large body of applied literature, have a tendency to just refer to ``mice'' as if that were one method. However, mice is a host of methods, depending on the regression that is chosen for imputation, from extremely successful method such \texttt{mice\_cart} to methods that only use simple regression and violate \textcolor{blue}{Rule 1} below.
\end{important}

\subsection*{Problem setup}

We try to only use mathematical symbols when they are actually helpful. Figure \ref{fig:illustrationfocond} summarizes our main notation; we assume there is a true underlying matrix $\mathbf{X}$ of dimension $n \times d$ with $n$ i.i.d. observations of vectors in $\mathbb{R}^d$. This is the unobserved data following a distribution $P$. What we actually observe is the matrices $\mathbf{M}$, and $\mathbf{X}^*$. The latter is the matrix of i.i.d. observations with missing values, arising from masking the observations in $\mathbf{X}$ by $\mathbf{M}$. That is, if $M_{i,j}=0$, then $X^*_{i,j}=X_{i,j}$ but if $M_{i,j}=0$, then $X^*_{i,j}=\textrm{NA}$, i.e. we do not observe $X_{i,j}$. We can thus observe $\mathbf{M}$ simply by checking with values were not observed (in \textsf{R} for instance, \texttt{M<- is.na(data) * 1}). Each observation (row) in $\mathbf{M}$ thus takes values in $\{0,1\}^d$. Each of these possible values is usually called a pattern. For instance, in Figure \ref{fig:illustrationfocond}, the first observation has pattern $(0,0,0)$ (fully observed) while the second one has pattern $(1,0,0)$ ($X_{1,2}$ missing). The well-known missingness mechanism--missing completely at random (MCAR), missing at random (MAR) and missing not at random (MNAR)-- usually framed in terms of the conditional distribution of $X \mid M$ (see e.g., \cite{RubinLittlebook,whatismeant}), can actually be phrased in terms of the following question: ``How can distributions change when moving from one pattern to another?''. We will not go into details about this here but refer to \cite{näf2025good}.

\begin{remark}
    Since this is a guide for applications, it should be read side by side with the associated code. As such, the code is available on \url{https://github.com/JeffNaef/Practical_Imputation_Guide} and will be referenced throughout the guide.
\end{remark}

\begin{figure*}[h!]
    \centering
\begin{tikzpicture}

\node at (0,0) {$\mathbf{X} = \begin{pmatrix}
x_{1,1} & x_{1,2} & x_{1,3} \\
x_{2,1} & x_{2,2} & x_{2,3} \\
x_{3,1} & x_{3,1} & x_{3,3}
\end{pmatrix}$};

\node at (4,0) {$\mathbf{M} = \begin{pmatrix}
0 & 0 & 0 \\
1 & 0 & 0 \\
1 & 1 & 0
\end{pmatrix}$};

\node at (8,0) {$\mathbf{X}^* = \begin{pmatrix}
x_{1,1} & x_{1,2} & x_{1,3} \\
\textrm{NA} & x_{2,2} & x_{2,3} \\
\textrm{NA} & \textrm{NA} & x_{3,3}
\end{pmatrix}$};

\end{tikzpicture}
    \caption{$\mathbf{X}$ is the assumed underlying full data, $\mathbf{M}$ is the vector of missing indicators and $\mathbf{X}^*$ arises when $\mathbf{M}$ is applied to $\mathbf{X}$. Thus each row of $\mathbf{X}$ (or $\mathbf{X}^*$) is an observation under a different pattern. Under missing completely at random (MCAR), no change is allowed. Under misisng at Random (MAR), the only constraint is that the distribution of $X_1, X_2 \mid X_3$ in the third pattern is the same as the unconditional one.}
    \label{fig:illustrationfocond}
\end{figure*}

\begin{longtable}{rlll}
  \hline
 & Methods & Languages & Implementations \\ 
  \hline
  1 & \texttt{knn} \citep{knnreference} & \textsf{R} & \texttt{bioconductor:impute} \citep{R_impute} \\ 
  2 & \texttt{mice\_cart} \citep{VANBUUREN2018} & \textsf{R} & \texttt{CRAN:mice} \citep{R_mice} \\ 
  3 & \texttt{mice\_drf} \citep{näf2025good} & \textsf{R} & \texttt{\href{https://github.com/KrystynaGrzesiak/miceDRF}{Git:KrystynaGrzesiak/miceDRF}} \\  
  4 & \texttt{mice\_norm} \citep{VANBUUREN2018} & \textsf{R} & \texttt{CRAN:mice} \citep{R_mice} \\ 
  5 & \texttt{mice\_norm\_nob} \citep{VANBUUREN2018} & \textsf{R} & \texttt{CRAN:mice} \citep{R_mice} \\ 
   6 & \texttt{mice\_norm\_predict} \citep{VANBUUREN2018} & \textsf{R} & \texttt{CRAN:mice} \citep{R_mice} \\  
  7 & \texttt{mice\_rf} \citep{VANBUUREN2018} & \textsf{R} & \texttt{CRAN:mice} \citep{R_mice} \\  
  8 & \texttt{missForest} \citep{stekhoven2012missforest} & \textsf{R} & \texttt{CRAN:missForest} \citep{R_missForest} \\ 
   \hline
\caption{List of imputation methods used in this guide. For the mice methods we use the naming convention of the \texttt{mice} \textsf{R} package} 
\label{tab:methods}
\end{longtable}

\section{The Three Properties of a Great Imputation}\label{Sec_1}

We start by loosely stating 3 ideal properties an imputation method should have in our view: 
\begin{rules}
The imputation should
    \begin{itemize}
    \item[Rule 1] be a distributional or stochastic method,
    \item[Rule 2] be highly flexible, ideally nonparametric,
    \item[Rule 3] work reasonably well under Missing at Random (MAR)\footnote{MAR loosely means that the probability of missingness only depends on what is observed, see \cite{whatismeant, näf2025good} for details.}.
\end{itemize}
\end{rules}

\paragraph*{\textcolor{blue}{Rule 3}} again reveals a certain bias, as our focus lies on the widely-used MAR condition originally introduced in \cite{Rubin_Inferenceandmissing}. There are also more recent alternative conditions that have their own merits, see e.g., \cite{directcompetitor0, näf2025good}. 

\paragraph*{\textcolor{blue}{Rule 2}} might seem self-explanatory, the more flexible a method the more likely it is to adapt to your data at hand. However, as with any estimation process, the more flexible the method, the more data points are required for accurate estimation. For less than say 200 observations a parametric imputation method like \texttt{mice.norm\_nob} might be better suited than a fully nonparametric method such as \texttt{mice\_cart}. This may also be true when $n$ is larger, but the data is high dimensional, i.e. $d \geq n$. Nonetheless in general, it seems desirable to have a method that is as flexible as possible.

\paragraph*{\textcolor{blue}{Rule 1}} is the most important of the three and it is the source of most pitfalls. We are not completely precise here, but with ``distributional'' or ``stochastic method'', we mean a method that draws imputations from a distribution instead of trying to predict the most likely value. As \cite[Chapter 2.6]{VANBUUREN2018} put it: Imputation is not prediction. In particular, stochastic methods allow for multiple imputation, because several distinct values can be drawn for each imputation. Imputation is thus a distributional estimation task; we do not know the missing values and it does not make sense trying to predict them. Instead, we want to impute such that we most closely reconstruct the original distribution.

\subsection*{Examples}

\subsubsection*{Gaussian Example}

\begin{figure}
    \centering
    \includegraphics[width=0.9 \linewidth]{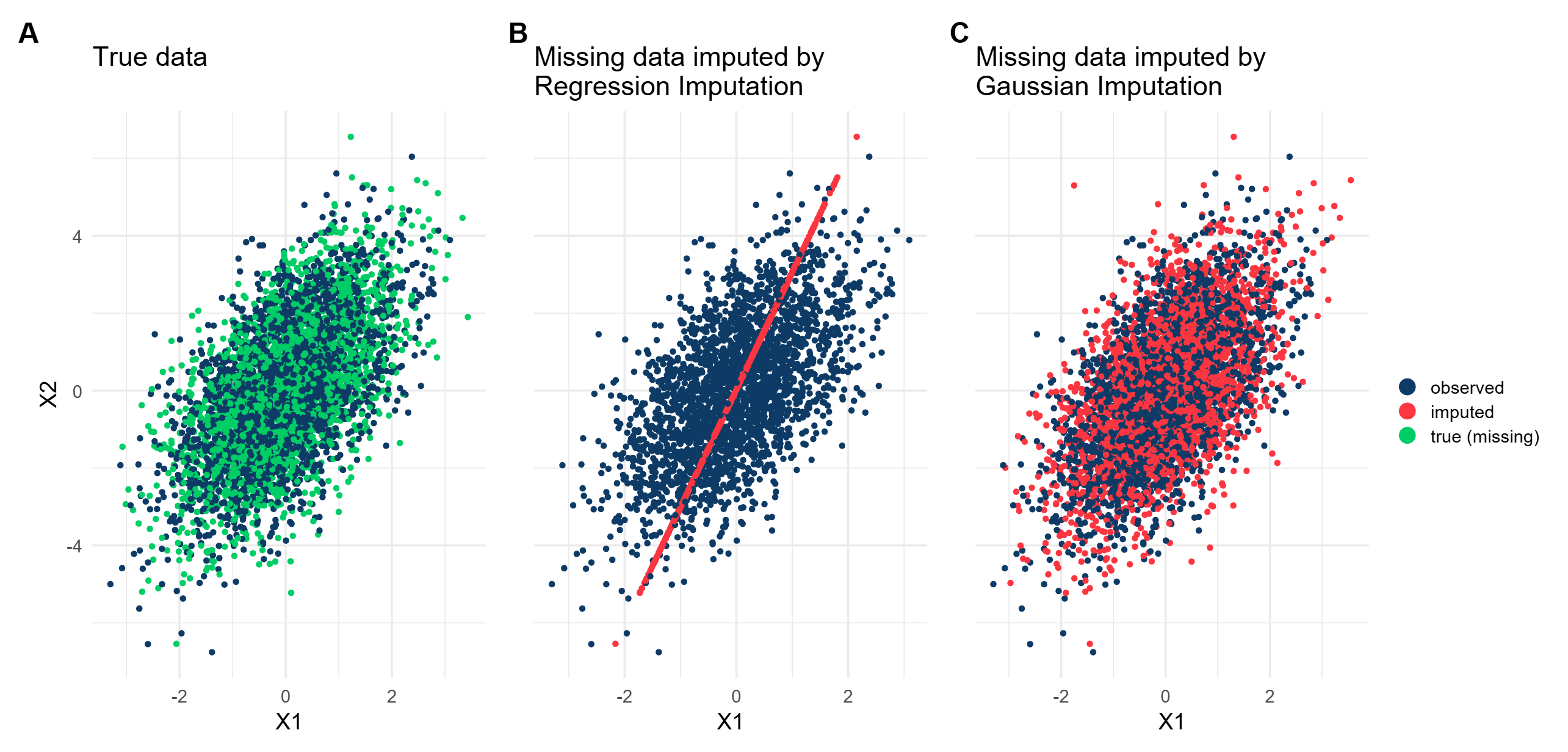}
    \caption{5000 observations of the bivariate Gaussian Example with around 50\% MCAR missing values in $X_1$.(A) Original observations (observed and missing), (B) Imputation by fitting a regression model and imputing the prediction (\texttt{mice\_norm.predict}), (C) Imputation by fitting a regression model and imputing by drawing from conditional Gaussian distribution (\texttt{mice\_norm}). Figure taken from \cite{OneBenchmarktorulethemall}.}
    \label{fig:MotivationExample}
\end{figure}

\begin{figure}
    \centering
      \includegraphics[width=1\linewidth]{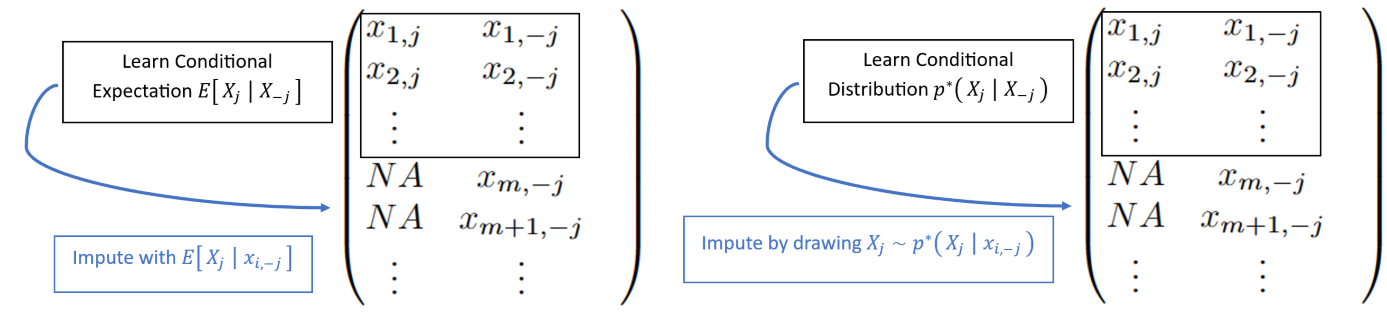}
    \caption{Imputation by learning from the part where $X_j$ is observed and imputing where it is missing, once using prediction (i.e. imputing by the estimated conditional expectation, the most likely value) on the left, and once using stochastic imputation (i.e. by drawing from the actually learned distribution) on the right.}
    \label{fig:imputationillustration}
\end{figure}

Figure \ref{fig:MotivationExample} taken from \cite{OneBenchmarktorulethemall} provides an immediate graphical illustration of this issue in a two dimensional Gaussian example. Essentially here we perform the mice algorithm of Figure \ref{fig:miceillustration}, with two different methods. The imputation on the very right (\texttt{mice\_norm}) fits a linear regression of $X_1$ onto $X_2$ in the pattern where both $X_1$ and $X_2$ are observed, leading to an estimate $\beta$ for the regression coefficient, and $\sigma^2$ for the residual variance. Instead, of directly trying to predict the missing values in the second pattern, using $\beta X_2$, the method draws the imputations from the distribution $N(\beta X_2, \sigma^2)$. This is also illustrated in Figure \ref{fig:imputationillustration} on the right. Compared to the the very left picture in Figure \ref{fig:MotivationExample}, we can see that this approach leads to an accurate recreation of the (unobserved) full distribution. Now compare this approach to the one that is often used instead (here represented by \texttt{mice\_norm.predict}, but the same would hold true for \texttt{missForest} or knn imputation). Here we try to predict the actual missing values by using $\beta X_2$ directly, giving a single value to impute.\footnote{One might argue that this is also a draw from a distribution, only that now the distribution trivially is a point mass around $\beta X_2$.} The detrimental effect of this is immediately visible in the middle picture, where the imputation results in a line through the point cloud. This is not just theoretical; the discrepancy can affect parameter estimation in downstream analyses. For instance, despite the high sample size, regressing $X_2$ onto the imputed $X_1$ leads to a bias for the regression imputation (estimated value $\approx$ 1.5, true value = 1). On the other hand, the \texttt{mice\_norm\_nob} imputation leads to an estimated value of $\approx$ 1.03, almost the same value as one obtains using the (unavailable) full data.

\begin{remark}
The Gaussian example is implemented in \texttt{Section\_2\_GaussianExample.R}, where the plot in Figure \ref{fig:MotivationExample} can also be created.
\end{remark}

\subsubsection*{Uniform Example}

  To further illustrate the point, we will now introduce a particularly tricky example that is designed to test if an imputation can truly handle MAR.  However, an imputation that imputes by prediction will also fail this test: 
\begin{uniformexample}\label{Example1}
 Consider variables $(X_1,X_2)$, each marginally distributed as a uniform between $[0,1]$, with $(X_1,X_2)$ being dependent, according to some copula. Moreover, we add $d-2$ uniform variables completely independent of the first two, simply because this tends to stabilize the more complex imputation methods a little bit. We then introduce missing values into this data set according to a mechanism designed to result in a complex MAR case. For details, see \cite[Example 6]{näf2025good}. The goal is then to (1) impute and (2) estimate the \emph{0.1-quantile} of $X_1$.  
\end{uniformexample}

\begin{figure}
    \centering
    \includegraphics[width=0.75\linewidth]{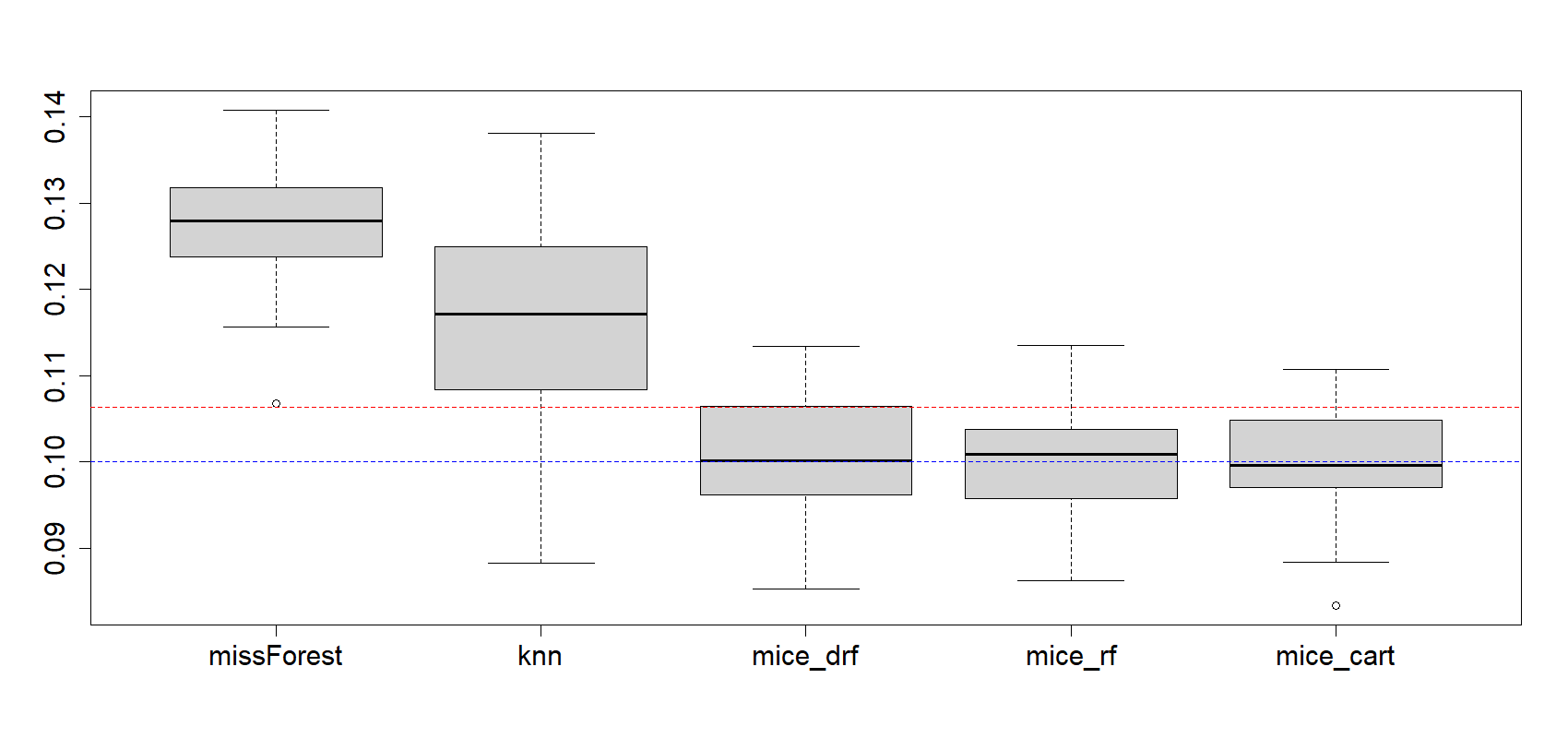}
    \caption{Result of the \textcolor{orange}{Uniform Example} over 50 simulations, tested on \texttt{missForest}, \texttt{knn}, \texttt{mice\_drf}, \texttt{mice\_rf}, \texttt{mice\_cart}. The vertical line represent the true value (0.1, in blue) and the true quantile of $X_1$ when considering only the observed values, without imputation ($-7 + \sqrt{49+15*0.1} \approx 0.106$, red). Methods are ordered by how closely the mean over the 50 replications is close to the true value of 0.1. See the file \texttt{Section\_2\_QuantileTest.R}.}
    \label{fig:QuantileTest}
\end{figure}

As mentioned this \textcolor{orange}{Uniform Example} is designed to do two things: First, it will quickly expose methods that do not actually work under MAR, as this is a rather tricky MAR example. Second, methods that do not meet \textcolor{blue}{Rule 1} will tend to have an even worse score. Figure \ref{fig:QuantileTest} showcases the second point for a few selected imputation methods over 50 simulation replications, $n=5000$, and $d=5$ (very low dimensional, but the problem is difficult enough). We note that the blue (lower) line corresponds to the true value of $\alpha=0.1$ (since $X_1$ is marginally uniform the desired 0.1 quantile is exactly 0.1). The upper read line instead shows the true (but biased) value we would get if we only estimated the quantile for $X_1$ when $X_1$ is observed, namely $-7 + \sqrt{49+15*0.1} \approx 0.106$.\footnote{This also is an important example showing why, in general, we cannot simply discard the missing values without biasing our analysis.}. The first 3 imputation methods, \texttt{mice\_drf}, \texttt{mice\_rf}, \texttt{mice\_cart}, approximately meet \textcolor{blue}{Rule 1} and do very well! However, missForest and knn imputation (\texttt{missForest}, \texttt{knn}) that do not meet \textcolor{blue}{Rule 1} and try to predict the missing values are over the red line and thus even worse than simply ignoring the missing data. Figure \ref{fig:QuantileTest} thus shows how trying to do the correct thing, i.e. not ignoring $X_1$ when it is missing, with the \emph{wrong} imputation method, can lead to \emph{worse} results.

\begin{remark}
The code for this experiment can be found on the GitHub link as \texttt{Section\_2\_QuantileTest.R}. It should be relatively straightforward to plug in any other method in the experiment, at least if it is implemented in \textsf{R} or can be used with Python and the \texttt{reticulate} package. Why not try your preferred method and see how it performs? If a method performs well in this experiment that would be interesting for this guide, as well as a good sign for the method.
\end{remark}

\section{How to choose the best Imputation Method}\label{Sec_2}

We have seen one way to assess how well an imputation method performs, by using a specifically designed simulated example. Of course this does not necessarily mean that a method that performs well here will also perform well on a given imputation task with a real-world data set, even if the missingness mechanism truly is Missing (Completely) at Random. 

As such, confronted with a data set with real missing values (where the true underlying data is not available), one might be tempted to follow the standard approach in practice: 

\begin{enumerate}
    \item delete a few randomly selected values in the data set that were originally observed,
    \item impute these artificially removed values, and
    \item compare the imputed value(s) to the true observed values using metrics such as RMSE (Root Mean Squared Error) or MAE (Mean Absolute Error).
\end{enumerate}
However, aside from the fact that deleting a value can render a proper MAR mechanism into an MNAR mechanism, this will once again favor methods that try to predict the missing values and violate \textcolor{blue}{Rule 1} above. This is because RMSE and MAE are optimized for the expectation and median respectively, so stochastic imputation methods, which introduce additional variance, may appear less attractive.

\subsection*{Imputation Scores}

Recently \cite{ImputationScores, näf2025Iscore} developed scoring methodologies for imputation methods, that should identify the best methods, similar as in the case of prediction. The latest version in \cite{näf2025Iscore}, essentially deletes observed values, imputes them and then checks the imputed vs the actually observed values, just as in the standard approach mentioned above. However, instead of using RMSE or MAE to compare one imputed value to one observed value, they impute multiple times, to have several imputed values for each test value. This brings us back to the realm of testing a sample from a predicted distribution (here the $N$ imputations) against a single test point (the previously deleted observed value) and there are convenient so-called \emph{proper scores} for this \citep{gneiting}. In particular, \cite{näf2025Iscore} use the energy score, thus naming the method ``energy-I-Score''. The score is implemented in the package \texttt{miceDRF}. Both the score itself and its implementation are not without its flaws, and the score can take a long time to evaluate. Improving this is a subject of further research.

\begin{remark}
    The website \url{https://krystynagrzesiak.github.io/miceDRF/articles/Example_IScore.html} offers a nice walkthrough of how to use the implementation of the energy-I-Score.
\end{remark}

We illustrate the use of this score in the \textcolor{orange}{Uniform Example} above. Getting the following result:

\begin{center}
\texttt{
\begin{tabular}{lrrrr}
mice\_drf & mice\_rf & mice\_cart & missForest & knn \\
0.1874387 & 0.1911953 & 0.2248830 & 0.2433831 & 0.2869977
\end{tabular}
}
\end{center}
Thus the score puts \texttt{mice\_drf} first, followed closely by \texttt{mice\_rf} and \texttt{mice\_cart} while placing the two methods that violate \textcolor{blue}{Rule 1} last. Note that the ordering of the first three methods is reversed compared to Figure \ref{fig:QuantileTest}. This is because their overall performance is very similar, and the score places strong emphasis on correct sampling (i.e. the ability to create multiple imputations). This is a finer point and we will not go into details here -- more importantly, choosing the method with the best score (\texttt{mice\_drf}) leads to great results in this example.

\begin{figure}
    \centering
    \includegraphics[width=1\linewidth]{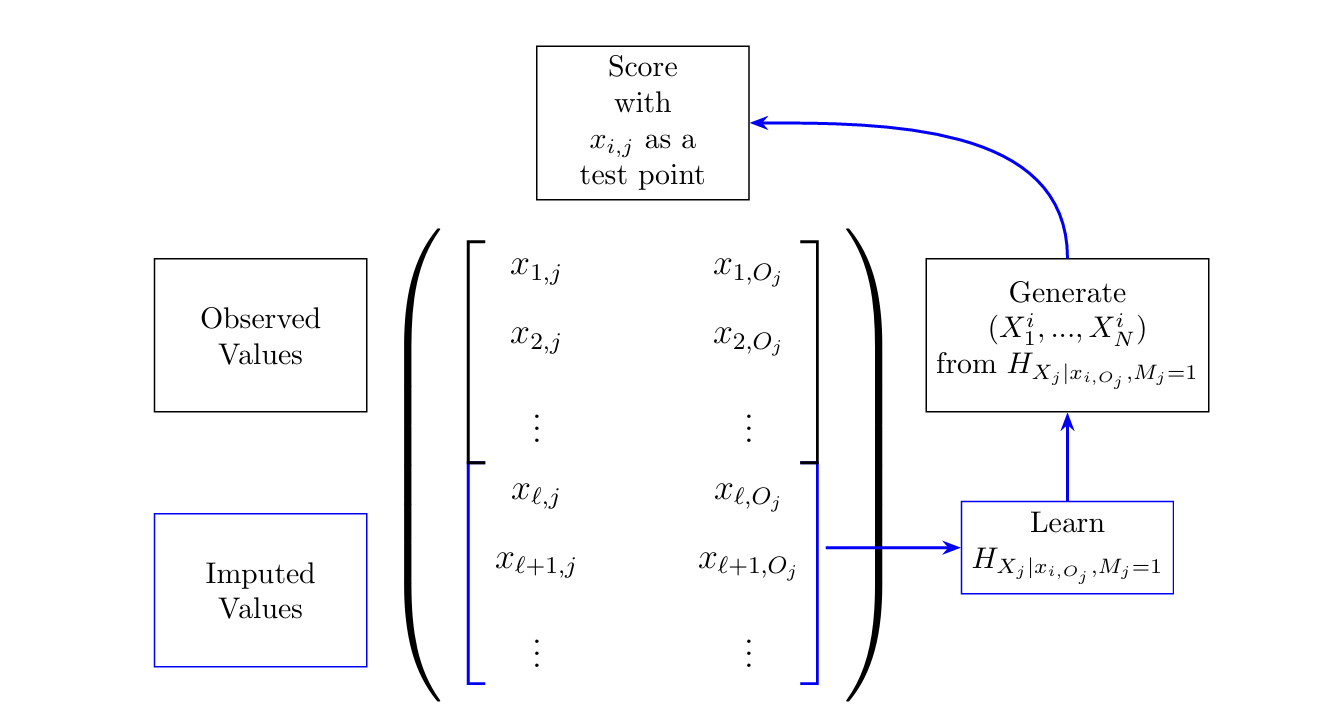}
    \caption{Illustration of the energy-I-Score, taken from \cite{näf2025Iscore}.}
    \label{fig:Scoreillustration}
\end{figure}

\begin{important}
        Most (modern) imputation methods are designed purely for numerical variables. In fact the benchmark \cite{OneBenchmarktorulethemall} found that only 21 out of 73 considered imputation methods were able to impute categorical variables, without making up new categories (i.e. imputing 2.5 when the categories were coded as 2 and 3.) This is clearly not realistic and another great advantage of \texttt{mice\_cart}, \texttt{mice\_rf} and \texttt{mice\_drf}. The score also has a version that works with mixed data (Categorical + Numerical data), see \url{https://krystynagrzesiak.github.io/miceDRF/articles/Example_IScore.html}. 
\end{important}

\section{Uncertainty Quantification}\label{Sec_3}

Traditionally, multiple imputation was combined with the so-called Rubin's rules to generate a variance estimate that would then be used to define confidence intervals for parameters (see e.g., \cite{VANBUUREN2018}). However, for (non-Bayesian) methods like the methods presented here, Rubin's rules tend underestimate the variance. This is usually framed as a shortcoming of the imputation method itself and arises from a Bayesian way of thinking where the parameter uncertainty of the imputation model itself is taken into account. As such, while this is a very elegant approach, there are two essential problems:
\begin{itemize}
    \item[(a)] Despite the strength of methods such as \texttt{mice\_cart} and \texttt{mice\_rf} and their ability to produce multiple imputation, using Rubin's rules might lead to an underestimation of the variance of a parameter and to too short confidence intervals.
    \item[(b)] Rubin's rules are not always straightforward to use, as they require an (asymptotic) expression of the variance of the estimator.
\end{itemize}
Would it thus be better to not use the powerful mice methods that appear to be some of the few methods achieving good results even in the \textcolor{orange}{Uniform Example}? We do not think so. Instead, we would like to use an approach that gives reliable (large sample) uncertainty estimation for imputation methods that use distributional imputation as described above.

One possible answer to this is the simple old bootstrap, which in fact was proposed a few times for parametric imputation, see e.g. \cite{Schomaker2018, Guan2024}. In particular, we 
\begin{itemize}
    \item[(0)] Calculate the main estimator as before, say $\hat{\theta}$, using the original data set (with imputation).
    \item[(1)] For $l=1,\ldots, L$ we shuffle the data each time (drawing by replacement), impute and then obtain a slightly different estimator, say $\theta^*_{l}$.
    \item[(2)] Use the normal approximation to build a confidence interval around $\hat{\theta}$:
    \[
    \hat{\theta} \pm q_{1-\alpha/2} \cdot \sigma^*,
    \]
    where $q_{1-\alpha/2}$ is the $1-\alpha/2$ quantile of a normal distribution (\texttt{qnorm(1-alpha/2)}) and $\sigma^*$ is the estimated standard deviation based on the bootstrap samples:
    \begin{align*}
        \sigma^*=\left( \sum_{l=1}^{L} (\hat{\theta} - \theta^*_{l})^2 \right)^{1/2}.
    \end{align*}
\end{itemize}

\begin{important}
    Note that there is no deep theory here that would justify the bootstrap approach and at best, this will give confidence intervals that are reliable for ``large'' sample sizes.
\end{important}

As far as we know, it was not extensively studied with the kind of nonparametric imputation we study here. We apply this procedure on the \textcolor{orange}{Uniform Example}, this time with $n \in \{1000, 2000\}$, but still for $d=5$. We take $L=30$ and impute with \texttt{knn}, \texttt{mice\_drf}, \texttt{mice\_rf}, \texttt{mice\_cart}. \texttt{missForest} is omitted here as it is relatively slow and will result in a very similar performance as \texttt{knn}. 

\begin{figure}[h]
    \centering
    \includegraphics[width=0.9\linewidth]{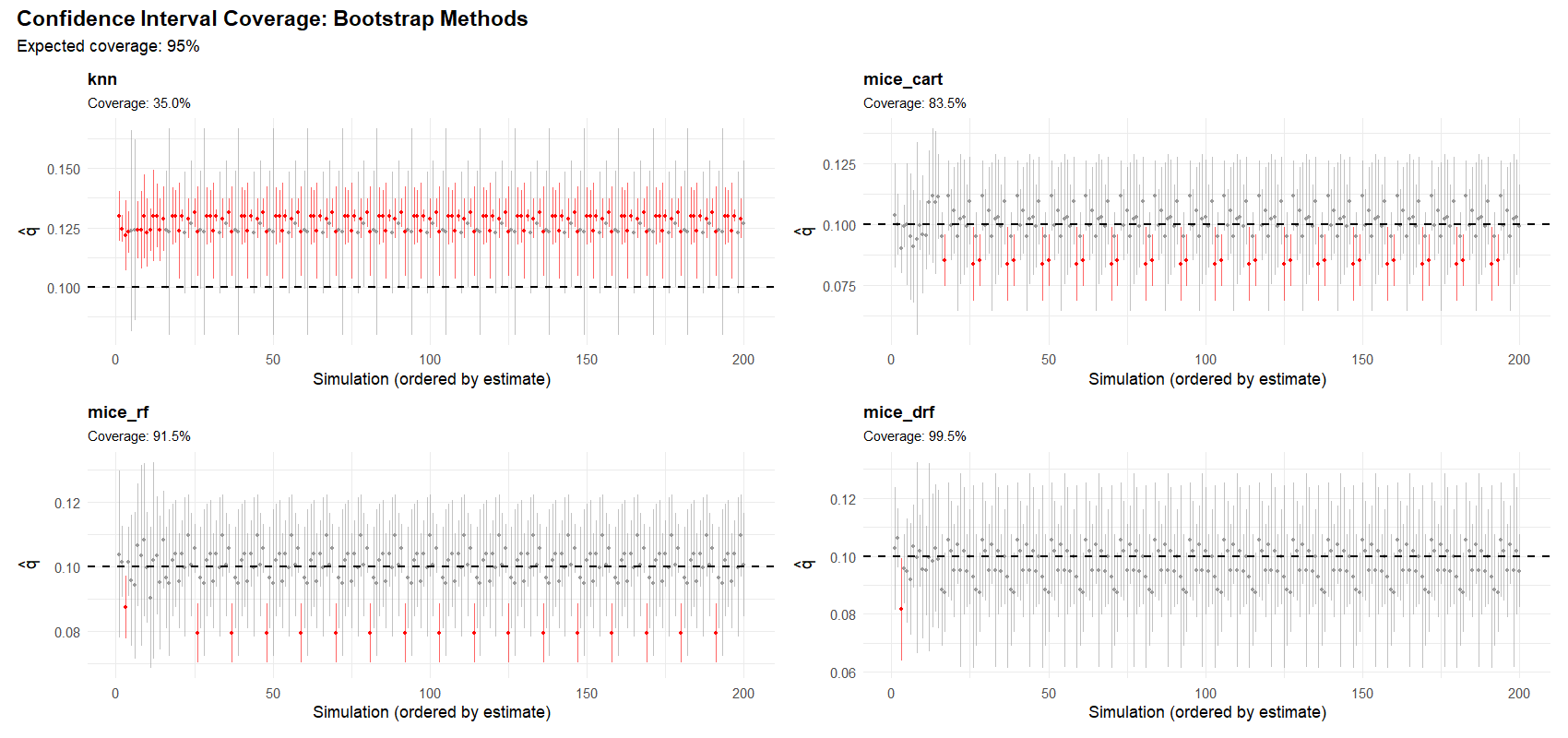}
    \caption{Coverage of the true value for the true value of 0.1 for $B=200$ replications, with $L=30$ bootstrap replications each time, $n=1000$, $d=5$. The normality approximation was used to build confidence intervals.}
    \label{fig:bootstrap1000}
\end{figure}

\begin{figure}[h]
    \centering
    \includegraphics[width=0.9\linewidth]{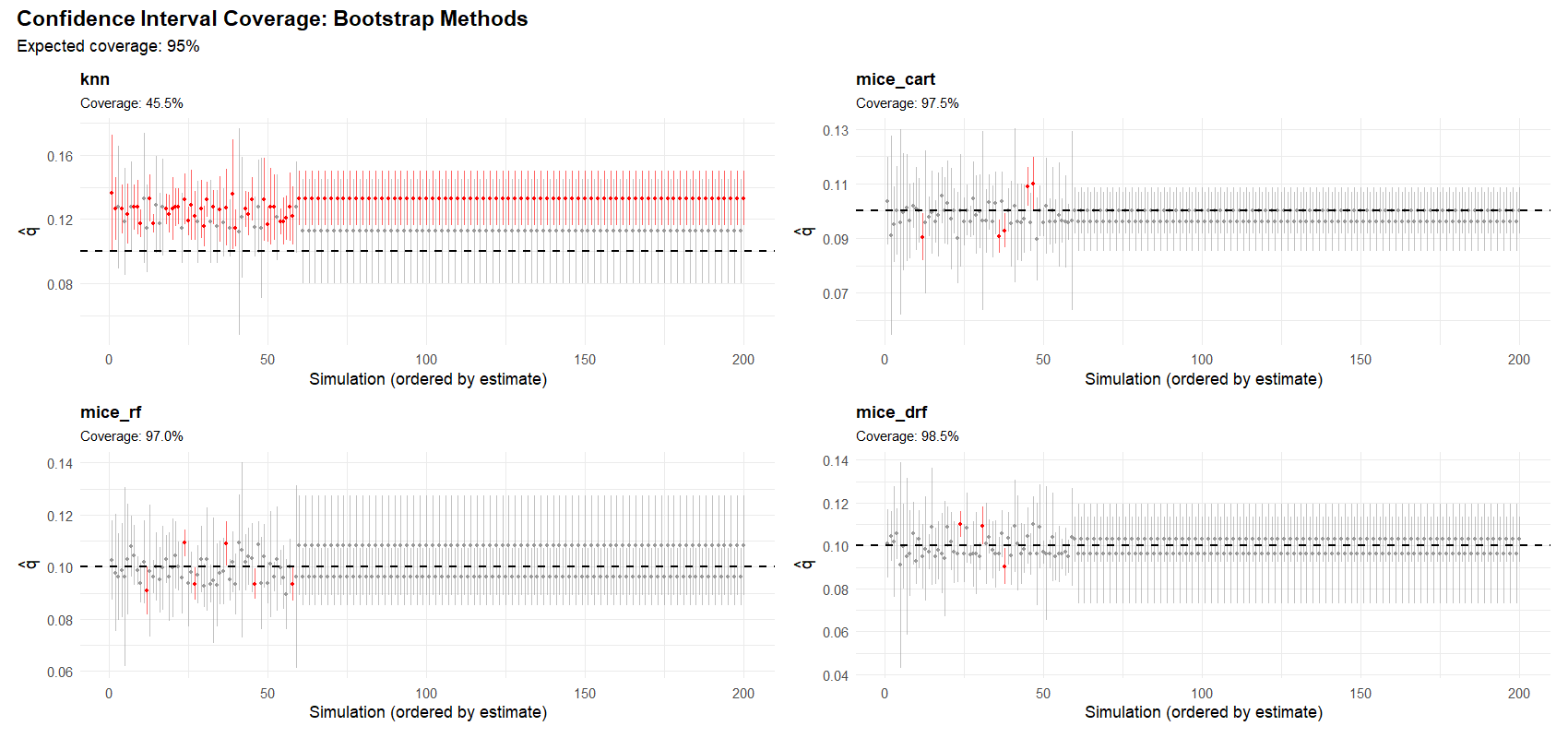}
    \caption{Coverage of the true value for the true value of 0.1 for $B=200$ replications, with $L=30$ bootstrap replications each time, $n=2000$, $d=5$. The normality approximation was used to build confidence intervals.}
    \label{fig:bootstrap}
\end{figure}

Figure \ref{fig:bootstrap1000} and \ref{fig:bootstrap} show the performance of this approach for the four imputation methods. The true value of $0.1$ is shown as the dashed horizontal line and for each of the $B=200$ simulations, a confidence interval is constructed. Ideally, we would like 95\% of these intervals to include the true value, while at the same time having intervals that are as short as possible. For the knn imputation, no matter how hard we try to build sensible intervals, its bias is simply too big and as a result it undercovers heavily (the true parameter is inside the confidence interval only 35\% and 45.5\% respectively). The picture would look almost identical for \texttt{missForest}, again underscoring the inadequacy of methods that fail \textcolor{blue}{Rule 1}. On the other hand, \texttt{mice\_rf}, \texttt{mice\_cart} are much better, though they still undercover for $n=1000$. Interestingly, \texttt{mice\_drf}, deemed the best method by the energy-I-Score for $n=5000$, has good coverage for $n=1000$, even if the produced intervals are maybe a bit too wide. For $n=2000$ on the other hand, all 3 distributional mice method show the desired coverage. These results might not seem very good given how simple the example is (only $d=5$ and uniform distributions), but we emphasize again that the MAR mechanism here is quite difficult.


\begin{remark}
    The code for this exercise can be found in the file \texttt{Section\_3\_Uncertainty.R}. It is a bit more complicated this time, but hopefully clear enough with the main text to replicate and use. Again, the reader is encouraged to plug-in and test their favorite imputation method.
\end{remark}

\begin{important}
    The computational properties of this procedure is rather horrible; the mice imputations considered here need to iterate several times over the dimensions of the data, and we even repeat this for $L=30$ times, which for high $d$ may take forever. On the other hand, with modern computing power this might not be such an issue. Nonetheless, faster (Rcpp or Python)-based implementations of \texttt{mice\_cart} and \texttt{mice\_rf} would be a great step forward. As amazing as the \texttt{mice} package is, it is mostly base \textsf{R} and thus rather slow.
\end{important}

\section{Conclusion}
The present paper attempts to give a short and concise, if biased, guide to modern imputation. We discussed that an imputation method should really try to recover the original data distribution (\textcolor{blue}{Rule 1}), how one might choose a good imputation method in practice, and how to deal with uncertainty quantification.

There are many questions still open, including what theoretical guarantees one actually can achieve and whether there is a more reliable way of uncertainty quantification for smaller samples. However, at least for small to mid-sized data sets, methods such as \texttt{mice\_cart} and \texttt{mice\_rf} show an incredible ability to recover the true data distribution under MAR.

\clearpage

\bibliographystyle{plain}
\bibliography{sample}

@article{Guan2024,
  author = {Guan, Qian and Yang, Shu},
  title = {A Unified Inference Framework for Multiple Imputation Using Martingales},
  journal = {Statistica Sinica},
  year = {2024},
  volume = {34},
  pages = {1649--1673}
}

@article{Schomaker2018,
  author = {Schomaker, M. and Heumann, C.},
  title = {Bootstrap Inference When Using Multiple Imputation},
  journal = {Statistics in Medicine},
  year = {2018},
  volume = {37},
  number = {14},
  pages = {2252--2266}
}

@article{OneBenchmarktorulethemall,
      title={Do we Need Dozens of Methods for Real World Missing Value Imputation?}, 
      author={Krystyna Grzesiak and Christophe Muller and Julie Josse and Jeffrey Näf},
      year={2025},
      eprint={2511.04833},
      journal={arXiv preprint arXiv:2511.04833},
      primaryClass={stat.CO}
}

@Manual{reticulate,
  title = {reticulate: Interface to 'Python'},
  author = {Kevin Ushey and JJ Allaire and Yuan Tang},
  year = {2025},
  note = {R package version 1.42.0,
    https://github.com/rstudio/reticulate},
  url = {https://rstudio.github.io/reticulate/},
}

@Article{benchmark_metabolomics1,
author={Kokla, Marietta
and Virtanen, Jyrki
and Kolehmainen, Marjukka
and Paananen, Jussi
and Hanhineva, Kati},
title={Random forest-based imputation outperforms other methods for imputing LC-MS metabolomics data: a comparative study},
journal={BMC Bioinformatics},
year={2019},
month={Oct},
day={11},
volume={20},
number={1},
pages={492}
}

@article{näf2025Iscore,
      title={How to rank imputation methods?}, 
      author={Jeffrey Näf and Krystyna Grzesiak and Erwan Scornet},
      year={2025},
      eprint={2507.11297},
      journal={arXiv preprint arXiv:2507.11297},
      primaryClass={stat.ME}
}

@article{näf2025good,
      title={What Is a Good Imputation Under MAR Missingness?}, 
      author={Jeffrey Näf and Erwan Scornet and Julie Josse},
      year={2026},
      eprint={2403.19196},
      journal={arXiv preprint arXiv:2403.19196},
      primaryClass={math.ST}
}

@ARTICLE{directcompetitor0,
  title    = "Multiple imputation for non-monotone missing not at random data
              using the no self-censoring model",
  author   = "Ren, Boyu and Lipsitz, Stuart R and Weiss, Roger D and
              Fitzmaurice, Garrett M",
  journal  = "Stat Methods Med Res",
  volume   =  32,
  number   =  10,
  pages    = "1973--1993",
  month    =  aug,
  year     =  2023,
  address  = "England",
  language = "en"
}

@article{knnreference,
    author = {Troyanskaya, Olga and Cantor, Michael and Sherlock, Gavin and Brown, Pat and Hastie, Trevor and Tibshirani, Robert and Botstein, David and Altman, Russ B.},
    title = "{Missing value estimation methods for DNA microarrays}",
    journal = {Bioinformatics},
    volume = {17},
    number = {6},
    pages = {520-525},
    year = {2001}
}

@article{benchmark_knn_adv1,
author = {Anil Jadhav, Dhanya Pramod and Krishnan Ramanathan},
title = {Comparison of Performance of Data Imputation Methods for Numeric Dataset},
journal = {Applied Artificial Intelligence},
volume = {33},
number = {10},
pages = {913-933},
year = {2019}
}

@article{benchmark_wang2022deep,
  title={Are deep learning models superior for missing data imputation in surveys? {E}vidence from an empirical comparison},
  author={Zhenhua Wang and Olanrewaju Akande and  Jason Poulos and Fan Li},
  journal={Survey Methodology},
  year={2022},
  volume={48},
  number={2}
}

@InProceedings{GAIN,
  title = 	 {{GAIN}: Missing Data Imputation using Generative Adversarial Nets},
  author =       {Yoon, Jinsung and Jordon, James and van der Schaar, Mihaela},
  booktitle = 	 {Proceedings of the 35th International Conference on Machine Learning},
  pages = 	 {5689--5698},
  year = 	 {2018}
}

@article{Rubin_Inferenceandmissing,
    author = {Rubin, Donald B.},
    title = "{Inference and missing data}",
    journal = {Biometrika},
    volume = {63},
    number = {3},
    pages = {581-592},
    year = {1976}
}

@Book{VANBUUREN2018,
author = {Stef {Van Buuren}}, 
title = {Flexible Imputation of Missing Data. Second Edition}, 
volume = {}, 
pages = {}, 
editor = {}, 
publisher = {Chapman \& Hall/CRC Press}, 
year = {2018}}

@article{whatismeant,
  title={{What is meant by “Missing at Random”?}},
  author={Seaman, Shaun and Galati, John and Jackson, Dan and Carlin, John},
  journal={Statistical Science},
  volume={28},
  number={2},
  pages={257-268},
  year={2013},
  publisher={Institute of Mathematical Statistics}
}

@article{gneiting,
author = {Tilmann Gneiting and Adrian E Raftery},
title = {Strictly Proper Scoring Rules, Prediction, and Estimation},
journal = {Journal of the American Statistical Association},
volume = {102},
number = {477},
pages = {359-378},
year  = {2007}

}

@book{RubinLittlebook,
author = {Roderick J. A. Little and Rubin B. Donald},
title = {Statistical Analysis with Missing Data},
year = {1986},
publisher = {John Wiley \& Sons, Inc.}
}

@article{mice,
    title = {{mice}: Multivariate Imputation by Chained Equations in {R}},
    author = {Stef {Van Buuren} and Karin Groothuis-Oudshoorn},
    journal = {Journal of Statistical Software},
    volume = {45},
    number = {3},
    year = {2011},
    pages = {1-67}
  }

@article{ImputationScores,
author = {Jeffrey Näf and Meta-Lina Spohn and Loris Michel and Nicolai Meinshausen},
title = {{Imputation scores}},
volume = {17},
journal = {The Annals of Applied Statistics},
number = {3},
publisher = {Institute of Mathematical Statistics},
pages = {2452 -- 2472},
year = {2023}
}

@article{benchmark_jager2021,
  title={A benchmark for data imputation methods},
  author={J{\"a}ger, Sebastian and Allhorn, Arndt and Bie{\ss}mann, Felix},
  journal={Frontiers in big Data},
  volume={4},
  pages={693674},
  year={2021},
  publisher={Frontiers}
}

@article{benchmark_joel2024performance,
  title={On the Performance of Imputation Techniques for Missing Values on Healthcare Datasets},
  author={Joel, Luke Oluwaseye and Doorsamy, Wesley and Paul, Babu Sena},
  journal={arXiv preprint arXiv:2403.14687},
  year={2024}
}

@article{benchmark_pereira2024imputation,
  title={Imputation of data Missing Not at Random: Artificial generation and benchmark analysis},
  author={Pereira, Ricardo Cardoso and Abreu, Pedro Henriques and Rodrigues, Pedro Pereira and Figueiredo, M{\'a}rio AT},
  journal={Expert Systems with Applications},
  volume={249},
  pages={123654},
  year={2024},
  publisher={Elsevier}
}

@article{benchmark_alam2023investigation,
  title={An investigation of the imputation techniques for missing values in ordinal data enhancing clustering and classification analysis validity},
  author={Alam, Shafiq and Ayub, Muhammad Sohaib and Arora, Sakshi and Khan, Muhammad Asad},
  journal={Decision Analytics Journal},
  volume={9},
  pages={100341},
  year={2023},
  publisher={Elsevier}
}

@article{benchmark_pavelchek2023imputation,
  title={Imputation of missing values for cochlear implant candidate audiometric data and potential applications},
  author={Pavelchek, Cole and Michelson, Andrew P and Walia, Amit and Ortmann, Amanda and Herzog, Jacques and Buchman, Craig A and Shew, Matthew A},
  journal={Plos one},
  volume={18},
  number={2},
  pages={e0281337},
  year={2023},
  publisher={Public Library of Science San Francisco, CA USA}
}

@article{benchmark_ge2023simulation,
  title={A simulation study on missing data imputation for dichotomous variables using statistical and machine learning methods},
  author={Ge, Yingfeng and Li, Zhiwei and Zhang, Jinxin},
  journal={Scientific Reports},
  volume={13},
  number={1},
  pages={9432},
  year={2023},
  publisher={Nature Publishing Group UK London}
}

@article{benchmark_seu2022intelligent,
  title={An intelligent missing data imputation techniques: A review},
  author={Seu, Kimseth and Kang, Mi-Sun and Lee, HwaMin},
  journal={JOIV: International Journal on Informatics Visualization},
  volume={6},
  number={1-2},
  pages={278--283},
  year={2022}
}

@article{benchmark_deforth2024performance,
  title={The performance of prognostic models depended on the choice of missing value imputation algorithm: a simulation study},
  author={Deforth, Manja and Heinze, Georg and Held, Ulrike},
  journal={Journal of Clinical Epidemiology},
  pages={111539},
  year={2024},
  publisher={Elsevier}
}

@article{benchmark_poulos2018missing,
  title={Missing data imputation for supervised learning},
  author={Poulos, Jason and Valle, Rafael},
  journal={Applied Artificial Intelligence},
  volume={32},
  number={2},
  pages={186--196},
  year={2018},
  publisher={Taylor \& Francis}
}

@article{benchmark_getz2023performance,
  title={Performance of multiple imputation using modern machine learning methods in electronic health records data},
  author={Getz, Kylie and Hubbard, Rebecca A and Linn, Kristin A},
  journal={Epidemiology},
  volume={34},
  number={2},
  pages={206--215},
  year={2023},
  publisher={LWW}
}

@article{benchmark_miao2022experimental,
  title={An experimental survey of missing data imputation algorithms},
  author={Miao, Xiaoye and Wu, Yangyang and Chen, Lu and Gao, Yunjun and Yin, Jianwei},
  journal={IEEE Transactions on Knowledge and Data Engineering},
  volume={35},
  number={7},
  pages={6630--6650},
  year={2022},
  publisher={IEEE}
}

@article{stekhoven2012missforest,
  title={MissForest—non-parametric missing value imputation for mixed-type data},
  author={Stekhoven, Daniel J and B{\"u}hlmann, Peter},
  journal={Bioinformatics},
  volume={28},
  number={1},
  pages={112--118},
  year={2012},
  publisher={Oxford University Press}
}

@Manual{R_impute,
title = {impute: impute: Imputation for microarray data},
author = {Trevor Hastie and Robert Tibshirani and Balasubramanian Narasimhan and Gilbert Chu},
year = {2024},
note = {R package version 1.80.0},
url = {https://bioconductor.org/packages/impute},
doi = {10.18129/B9.bioc.impute},
}

@Manual{R_mice,
title = {mice: Multivariate Imputation by Chained Equations},
author = {Stef van Buuren and Karin Groothuis-Oudshoorn and Gerko Vink and Rianne Schouten and Alexander Robitzsch and Patrick Rockenschaub and Lisa Doove and Shahab Jolani and Margarita Moreno-Betancur and Ian White and Philipp Gaffert and Florian Meinfelder and Bernie Gray and Vincent Arel-Bundock and Mingyang Cai and Thom Volker and Edoardo Costantini and Caspar van Lissa and Hanne Oberman and Stephen Wade},
year = {2023},
note = {R package version 3.16.0}
}

@Manual{R_missForest,
title = {missForest: Nonparametric Missing Value Imputation using Random Forest},
author = {Daniel J. Stekhoven},
year = {2022},
note = {R package version 1.5}
}

@article{benchmark_wongkamthong2023comparative,
  title={A comparative study of imputation methods for multivariate ordinal data},
  author={Wongkamthong, Chayut and Akande, Olanrewaju},
  journal={Journal of Survey Statistics and Methodology},
  volume={11},
  number={1},
  pages={189--212},
  year={2023},
  publisher={Oxford University Press}
}

\end{document}